\documentclass[preprint,authoryear,12pt]{elsarticle}

\usepackage[dvips]{color}

\usepackage{natbib}
\usepackage{setspace}
\usepackage{amsmath}
\usepackage{graphicx}
\usepackage{etoolbox}
\makeatletter
\patchcmd{\Ginclude@eps}{"#1"}{#1}{}{}
\makeatother

\usepackage{verbatim}
\usepackage{rotating}
\usepackage{multirow}
\usepackage{longtable}
\usepackage{booktabs}
\usepackage{subfigure}
\usepackage{lineno}

\usepackage{fancyhdr}
\begin{document}
\begin{frontmatter}

\title{A maximum likelihood estimate of natural mortality for brown tiger prawn ({\it Penaeus esculentus}) in Moreton Bay (Australia)}

\author[label1,label2]{Marco Kienzle\corref{cor1}}
\ead{Marco.Kienzle@daff.qld.gov.au}
\author[label3]{David Sterling}
\ead{djstgs@bigpond.com}
\author[label4]{Shijie Zhou}
\ead{shijie.zhou@csiro.au}
\author[label5]{You-Gan Wang}
\ead{you-gan.wang@uq.edu.au}
\address[label1]{Queensland Department of Agriculture, Fisheries and Forestry\\Ecosciences Precinct, Joe Baker St, Dutton Park, Brisbane, QLD 4102, Australia}
\address[label2]{University of Queensland, School of Agriculture and Food Sciences, St. Lucia, QLD 4072, Australia}
\address[label3]{Sterling Trawl Gear Services, Manly, QLD, Australia}
\address[label4]{CSIRO Marine and Atmospheric Research and Wealth from Oceans Flagship, PO Box 2583, Brisbane, QLD 4001, Australia}
\address[label5]{Centre for Applications in Natural Resource Mathematics, School of Mathematics and Physics, University of Queensland}

\cortext[cor1]{Corresponding author. Tel.: +61 (0)7 3255 4232; fax: +61 (0)7 3844 8235. E-mail address: Marco.Kienzle@daff.qld.gov.au}

\begin{abstract}
The delay difference model was implemented to fit 21 years of brown tiger prawn ({\it Penaeus esculentus}) catch in Moreton Bay by maximum likelihood to assess the status of this stock. Monte Carlo simulations testing of the stock assessment software coded in C++ showed that the model could estimate simultaneously natural mortality in addition to catchability, recruitment and initial biomasses. Applied to logbooks data collected from 1990 to 2010, this implementation of the delay difference provided for the first time an estimate of natural mortality for brown tiger prawn in Moreton Bay, equal to $0.031 \pm 0.002$ week$^{-1}$. This estimate is approximately 30\% lower than the value of natural mortality (0.045 week$^{-1}$)  used in previous stock assessments of this species.

\end{abstract}

\begin{keyword}
tiger prawn \sep natural mortality \sep eco-physiological response
\end{keyword}

\end{frontmatter}

%\linenumbers

\section{Introduction} %[ Why are natural mortality rate estimates important ? ]

In fisheries science, estimates of mortality rates are essential to evaluate the status of exploited marine resources. The magnitude of natural mortality affects our predictions of stock productivity, making this parameter one of the most important to fisheries management \citep{lee11a}. Natural mortality is notoriously difficult to estimate from catch and effort data \citep{Wang99a, zhou11a}. Contemporary fisheries science approaches advocate estimation of natural mortality within stock assessment models using catch data to separate fishing mortality from estimates of total mortality instead of using inaccurate indirect methods \citep{Maunder25022014}. In practice, many researchers are cautious about estimating this population dynamic parameter using stock assessments because model mis-specifications and poor data quality have often led to un-realistic estimates \citep{Francis2012LTE}. As a consequence, the standard practice is to use an externally estimated value of natural mortality as a fixed input parameter to fit stock assessment models. \\

The stock assessment of brown tiger prawn ({\it Penaeus esculentus}) in Moreton Bay \citep{Kienzle2014138} applied a standard approach by borrowing natural mortality estimates from the Northern Prawn Fishery \citep{dichmont2003application}. The value of natural mortality (0.045 week$^{-1}$) estimated by \cite{Wang99a} using a depletion method has been widely used for over four decades without validation. Extending this approach to combine estimates of catchability and natural mortality for the Northern Prawn Fishery (NPF), \citep{Zhou25022014} reached a substantially lower value for natural mortality (approx. 0.03 week$^{-1}$) suggesting natural mortality might not have been as high as previously thought and providing some incentive for further research into this topic. The logic of \cite{Wang99a}'s approach was to subset catch and effort data using the period when recruitment could be assumed to be negligible. But it was based on a single year of data from a single stock region and therefore lacked generality. This method would have required substantial modifications before being applicable to Moreton Bay fishery to account for targeted effort and temperature-driven variations of availability, two processes identified as having major influences on this fishery \citep{Kienzle2014138}. The cost and effort of customizing Wang's method to estimate natural mortality for tiger prawn in Moreton and possibly integrate it to the stock assessment to account for negative correlations between fishing and natural mortality might have been greater, if at all possible, than developing the stock assessment itself. \\

Therefore, in this paper, we propose to use a different approach to the problem of estimating natural mortality by maximum likelihood for tiger prawn in Moreton Bay. The delay difference model \citep{sch85a, Der80a} used for stock assessment was substantially enhanced since \cite{Kienzle2014138} to account for additional sources of known variability in this fishery, including environmental factors, that improved the predictions of observed variations of catch from 1990 to 2010. The delay difference model, implemented in C++, was submitted to an extensive set of tests to check the quality of this program and ensure that it performed according to expectations \citep{stroustrup2009programming}. These tests involving simulations of datasets over parameters ranges consistent with the biology of this species revealed that the model could provide an accurate estimate of simulated natural mortality as well as all the other parameters required for stock assessment, excluding those pertaining to somatic growth. Therefore we estimated natural mortality using real data which provided a realistic value for tiger prawn and suggested to use this approach in future assessments. \\

\section{Materials and methods} 

   \subsection{Stock assessment model} \subsubsection{Population dynamics}

A Schnute-Deriso delay-difference model \citep{sch85a, Der80a, hil92b} was used to estimate weekly variations in biomass ($B_{t}$) of brown tiger prawns in Moreton Bay between 1990 and 2010 

\begin{equation}
B_{t} = s_{t-1} \ B_{t-1} + \rho \ s_{t-1} \ B_{t-1} - \rho \ s_{t-1} \ s_{t-2} \ B_{t-2} - s_{t-1} \ \rho \ w_{k-1} \ R_{t-1} + w_{k} \ R_{t} \ ,\  3 \leq t \leq 1092
\end{equation}

Sex-combined growth parameters ($\rho$, $w_{k-1}$ and $w_{k}$), derived from von Bertalanffy estimates \citep{grib94a}, were fixed in the model (Tab.~\ref{tab:ModelPar}). This model assumed all prawns were fully recruited to the fishery (knife-edged selectivity) at an age of 22 weeks ($k=22$), weighing 19.5 grams. $w_{k-1}$, the pre-recruitment weight was interpreted as a parameter rather than the actual weight at age $k-1$ and was estimated according to \cite{sch85a}. Survival ($s_{t}$) varied as a function of a fixed natural mortality rate ($M$, Tab.~\ref{tab:ModelPar}) and fishing mortality ($F_{t}$) proportional to effort ($F_{t} = q E_{t}$)

\begin{equation}
s_{t} = \exp[- (M + q E_{t})]
\end{equation}

\noindent where catchability ($q$) was estimated.\\

This model estimated the magnitude of recruitment in each week ($R_{t}$) using 1 parameter to describe total recruitment in each 21 years ($R_{y}$) between 1990 and 2010 and 2--22 parameters (constant and year specific $\mu$, $\kappa$) from the von Mises probability density function \citep{mardia1999directional} to allocate a proportion of the total recruitment within each year to each 52 weeks

\begin{equation}
f(x|\mu, \kappa) = \frac{\exp[\kappa \ cos(x-\mu)]}{2 \pi I_{0}(\kappa)}
\end{equation}

\noindent where $I_{0}(x)$ is the modified Bessel function of order 0.\\

   \subsection{Monte Carlo simulations} The program that simulated fisheries data was originally developed for quality control purposes, to ensure that the stock assessment software coded in C++ \citep{Kienzle2014138} could be relied on to do what it was supposed to do with an acceptable error rate \citep{stroustrup2009programming}. It was designed with Moreton Bay brown tiger prawn fishery in mind, drawing parameters values from ranges consistent with this stock (Tab.~\ref{tab:SimulationParameters.tex}). The dynamics of the fishery was simulated with an age-structured model which alleviated the criticism raised by \cite{Francis2012LTE} about validity of simulation testing using the same model to generate datasets and estimate parameters. Age ($a$) varied between 22 and 156 weeks ($ 1 \leq a \leq p, p= 135 $). Each simulated dataset consisted of 4 years of data at weekly timesteps ( $1 \leq t \leq n, n= 4 \times 52$).\\

Individual's weight at age $a \ge k$ (where $k=22$ weeks is the age at recruitment to the fishery) was calculated according to \cite{sch85a} Eq. 1.14:
\begin{equation}
W_{t,a} = w_{t-1,k-1} + (w_{t,k} -  w_{t-1,k-1} ) \frac{1-\rho^{1-a-k}}{1-\rho}  
\end{equation}

where $w_{t-1,k-1}$, $w_{t,k}$ and $\rho$ where parameters fixed to 17.756, 19.5 and 0.96328 to represent brown tiger prawn somatic growth. Weight at age were kept constant between time steps and represented using a matrix ($W$) with dimension $n \times p$ having identical rows.

The value of several parameters were generated at random including catchability, natural mortality, von Mises parameters for the recruitment distribution, magnitude of recruitment in each year (Tab.~\ref{tab:SimulationParameters.tex}). The matrix of effort ($E$ with dimension $n \times p$) had identical columns made of a vector of effort ($E_{t}$) randomly generated from observed weekly effort in Moreton Bay fishery. $F$ is the fishing mortality matrix defined as $F=qE$ where $q$ is a randomly generated scalar representing the catchability (Tab.~\ref{tab:SimulationParameters.tex}). $M$ is a constant matrix with dimension $n \times p$. Natural mortality was generated using a uniform distribution varying between 0.025 and 0.065 week$^{-1}$.

Recruitment to the population ($N_{.,1}$) was generated using the product of a random yearly magnitude of recruitment (Tab.~\ref{tab:SimulationParameters.tex}) and distributing across 52 weeks using a von Mises distribution with randomly generated mean and standard deviation (Tab.~\ref{tab:SimulationParameters.tex}). 

The matrix of number at age in the population ($N$) was filled simulating the dynamic of $n+p-1$ cohorts, applying the declining in abundance along each cohort:

\begin{equation}
N_{i+1,j+1} = N_{i, j} \times {\rm exp} ( - (M_{i,j} + q E_{i,j})), {\rm for} 1 \leq i \leq n-1 \ {\rm and} \ 1 \leq j \leq p-1
\end{equation}

Hence the matrix of population biomass ($B$) was calculated using:
\begin{equation}
B = N * W
\end{equation}

And the vector of biomass for each timestep:

\begin{equation}
B_{t} = \sum_{a} B
\end{equation}

The matrix of catch at age ($C$, with dimension $n \times p$) was calculated using Baranov's catch equation \citep{quin99b} 

\begin{equation}
C = \frac{F}{F+M} B (1 - {\rm exp}(-(F+M))
\end{equation}

Finally the time series of catch were calculating summing across columns the following expression

\begin{equation}
C_{t} = \sum_{a} C
\end{equation}

The time series of catch and effort were altered by "white noise" multiplying each element by a random factor uniformly distributed between 0.9 and 1.1 to introduce $\pm$ 10\% error into the dataset passed to the stock assessment. 

This simulator was written in R \citep{R}. The code is available from M. Kienzle.

   \subsection{Tiger prawns in Moreton Bay}

%          \subsubsection{Water temperature data} \input{../TowardAnEcologicalStockAssessmentModel/TemperatureData.tex}
          \subsubsection{Water temperature data} 
\begin{figure}[h!]
     \begin{center}
       \includegraphics[scale=0.6, angle = 0]{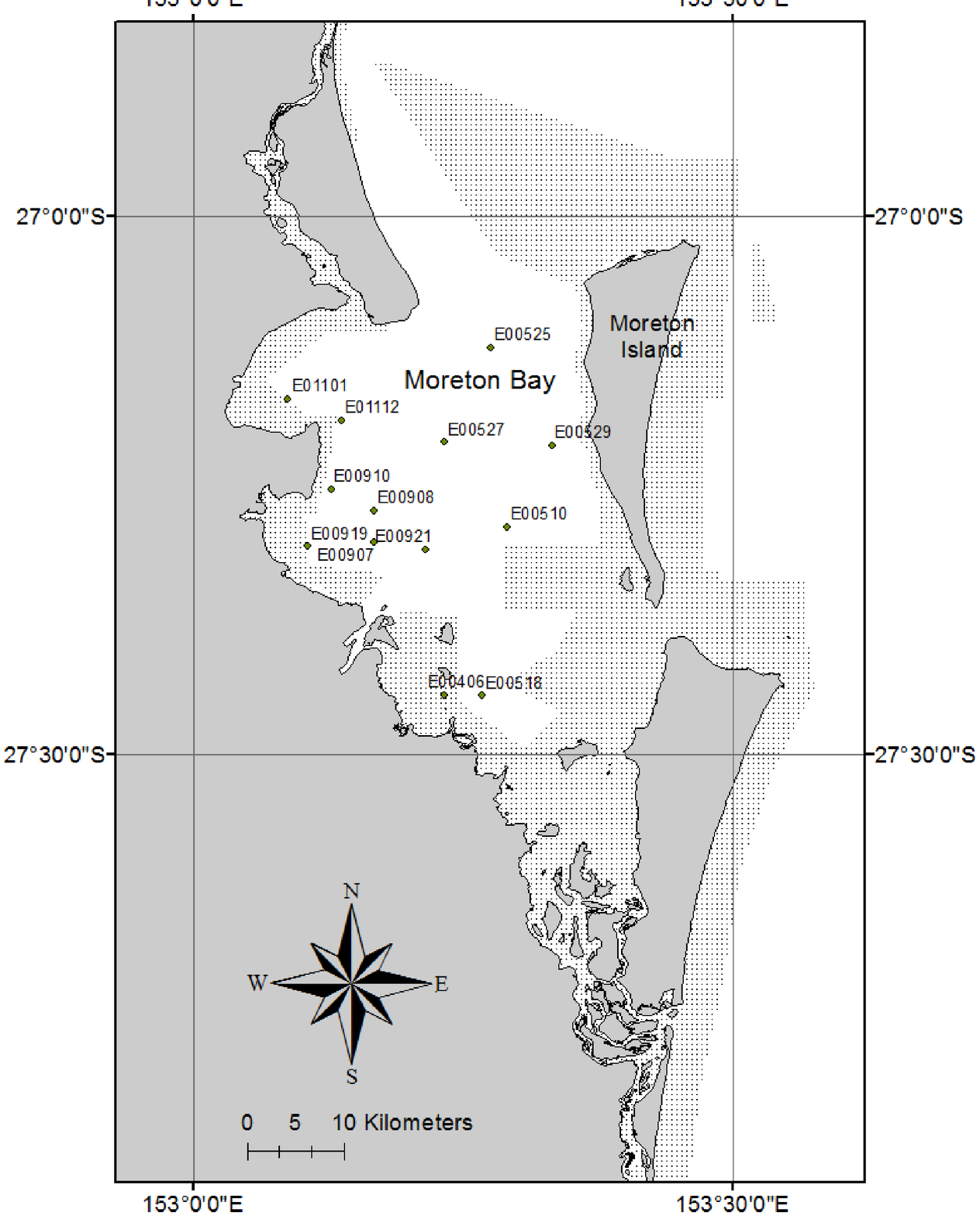}
       \caption{Location of the sampling site in Moreton Bay.}
       \label{fig:MapsOfSamplingSite}
     \end{center}
  \end{figure}

Moreton Bay is a shallow body of sea water, with an average depth of 6.8 m and a maximum of 36 m \citep{nla.cat-vn5267095}. Since 2002, water temperature has been measured monthly at 85 stations spread across the Bay by the Ecosystem Health Monitoring Program for the Healthy Waterways Program of South East Queensland (http://healthywaterways.org). Station E00527 is located close to the center of the brown tiger prawn trawling grounds (Fig.~\ref{fig:MapsOfSamplingSite}). Its bottom temperature measurements were used to quantify the emergence behaviour of this benthic specie. Maximum daily air temperature were recorded continuously by the Australian Bureau of Meteorology (BOM, \cite{BOM}) at Cape Moreton Lighthouse, station 040043. A linear regression between these two time series of measurements was used to estimate daily bottom temperature on the fishing ground.

%          \subsubsection{Duration of emergence} \input{../TowardAnEcologicalStockAssessmentModel/SigmoidModelOfHillExperimentalData.tex}
          \subsubsection{Duration of emergence} \cite{hil85a} measured experimentally the effect of temperature ($T$) on the duration of emergence ($y$) of brown tiger prawn. This author assumed a linear relationship between these variables whereas he used a sigmoid function for this relationship in crab {\it Scylla serrata} \citep{hill80a}. Biologically, an asymptotic relationship makes more sense as nocturnal emergence time cannot increase infinitely with time. As a consequence, a sigmoid function was fitted to \cite{hil85a}'s experimental data
  
\begin{equation}
y(T) = \frac{a}{1  + {\rm exp}( b - c T)}
\end{equation} 

in place of a linear relationship using the non-linear regression functions in R \citep{R}. Goodness of fit was assessed with a $\chi^{2}$ statistics \citep{bra98b}. An index of availability ($\gamma(T)$ varying between 0 and 1 at a given temperature, in $^{o}$C, was calculated using the predicted duration of emergence ($\hat{y}(T)$) divided by the asymptote of the sigmoid model

\begin{equation}
\gamma(T) = \frac{\hat{y}(T)}{a}
\end{equation}

This index of availability was used as a multiplier to effort in the stock assessment model \citep{Kienzle2014138} to test the assumption that emergence time affected catchability.

          \subsubsection{Refining the model to estimate natural mortality} The present work extended the delay difference model presented in \cite{Kienzle2014138}. For consistency, model numbers were kept identical and the reader is referred to the previous study for details on model 1 to 7. \\

The first area investigated concerned the prawn emergence model: model 8 replaced the linear model in model 4 by a sigmoid relationship. Model 9 used estimated daily water temperature rather than averaged monthly temperature off Cape Moreton. Model 10 was identical to model 9 but estimated natural mortality rather than fixing it to 0.045 week$^{-1}$. Finally model 11 allowed for timing of recruitment to vary from year to year.\\

\section{Results} 

   \subsection{Monte Carlo simulations} 
\begin{figure}[!ht]
 \subfigure[]{ % FROM THE SUBFIGURE PACKAGE
    \label{fig:a}
    \begin{minipage}[b]{0.5\textwidth}
     \includegraphics[scale=0.3, angle = -90]{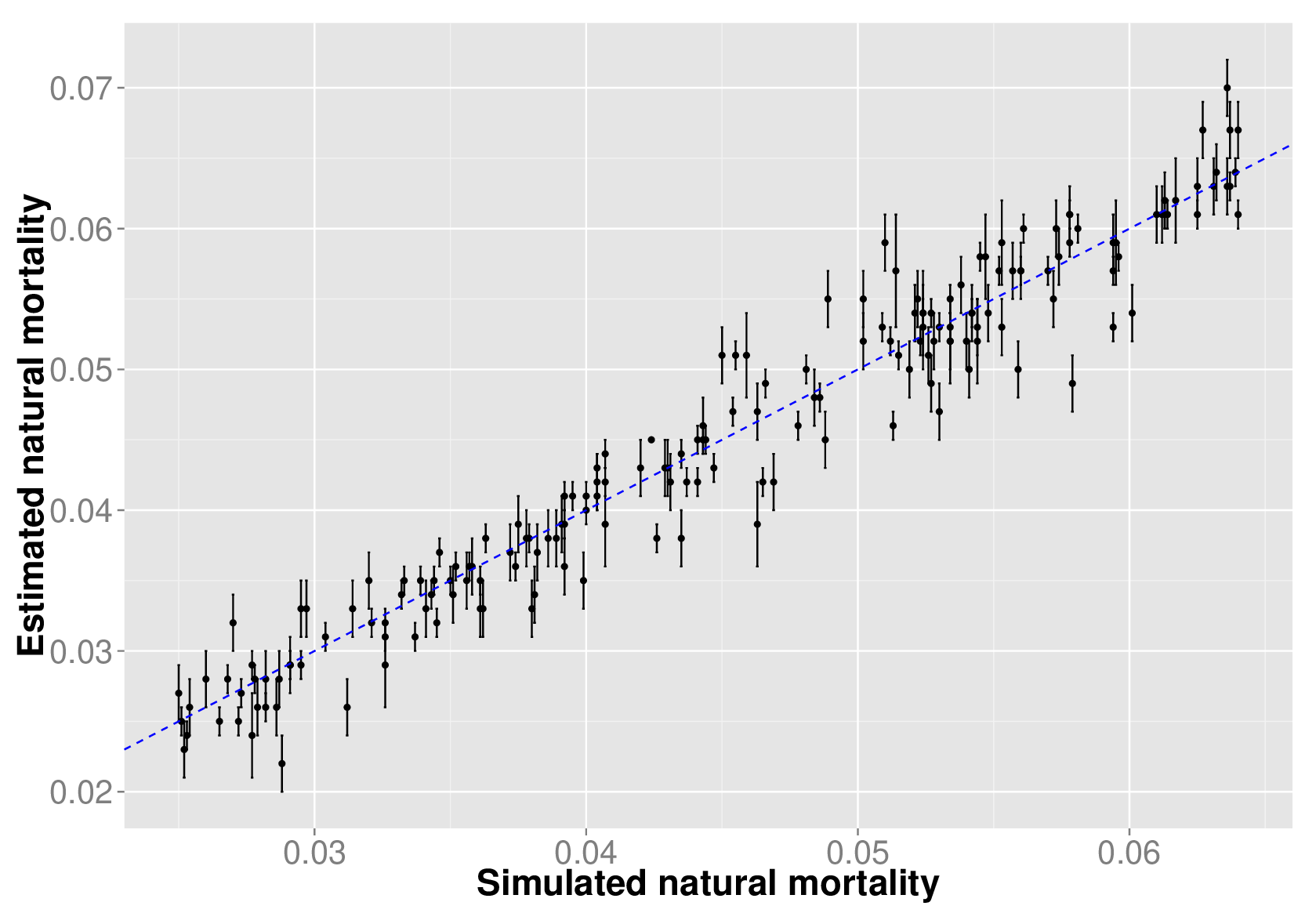}
   \end{minipage}
}
 \subfigure[]{ % FROM THE SUBFIGURE PACKAGE
    \label{fig:b}
    \begin{minipage}[b]{0.5\textwidth}
    \includegraphics[scale=0.3, angle = -90]{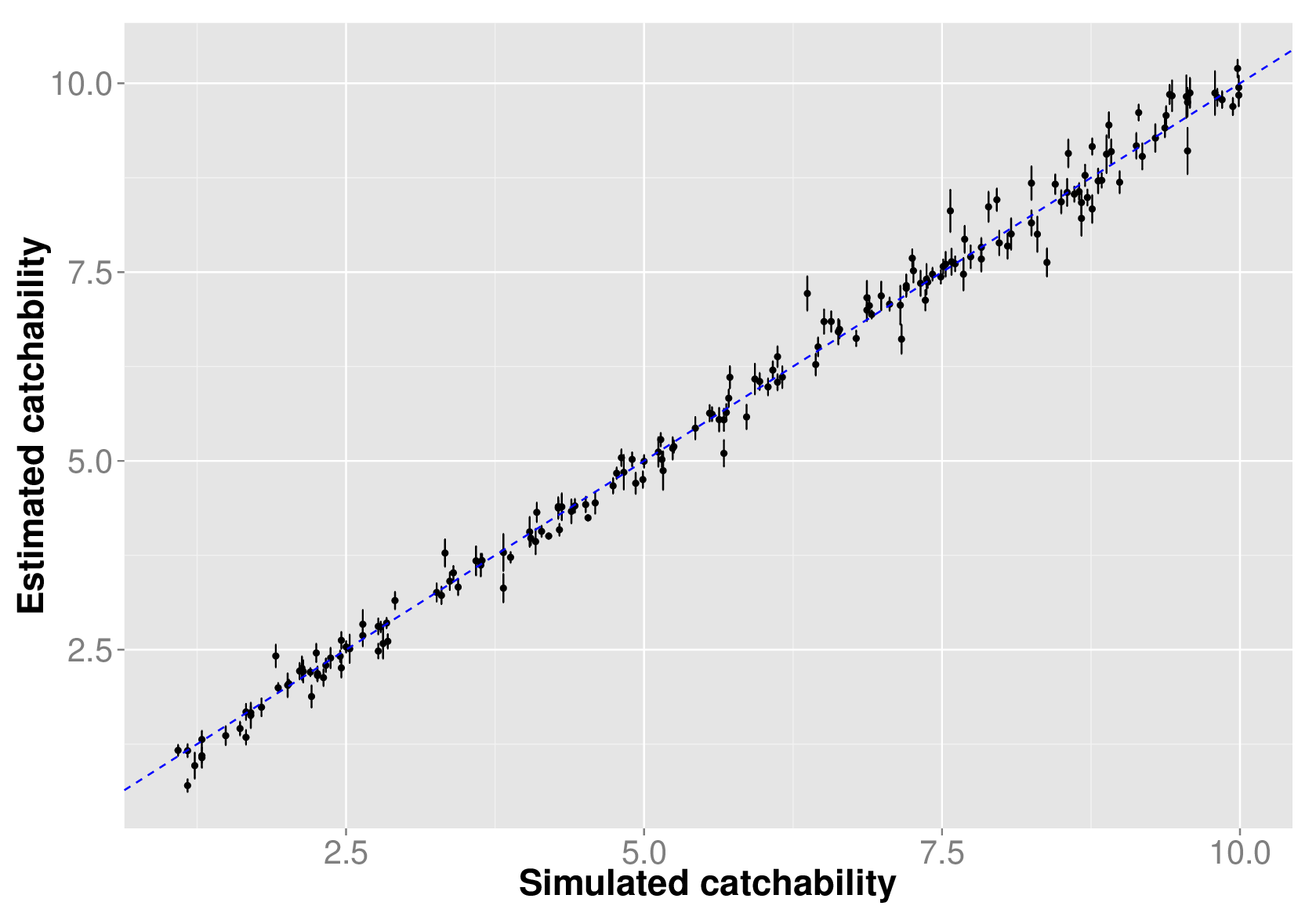}
   \end{minipage}
}
 \subfigure[]{ % FROM THE SUBFIGURE PACKAGE
    \label{fig:c}
    \begin{minipage}[b]{0.5\textwidth}
    \includegraphics[scale=0.3, angle = -90]{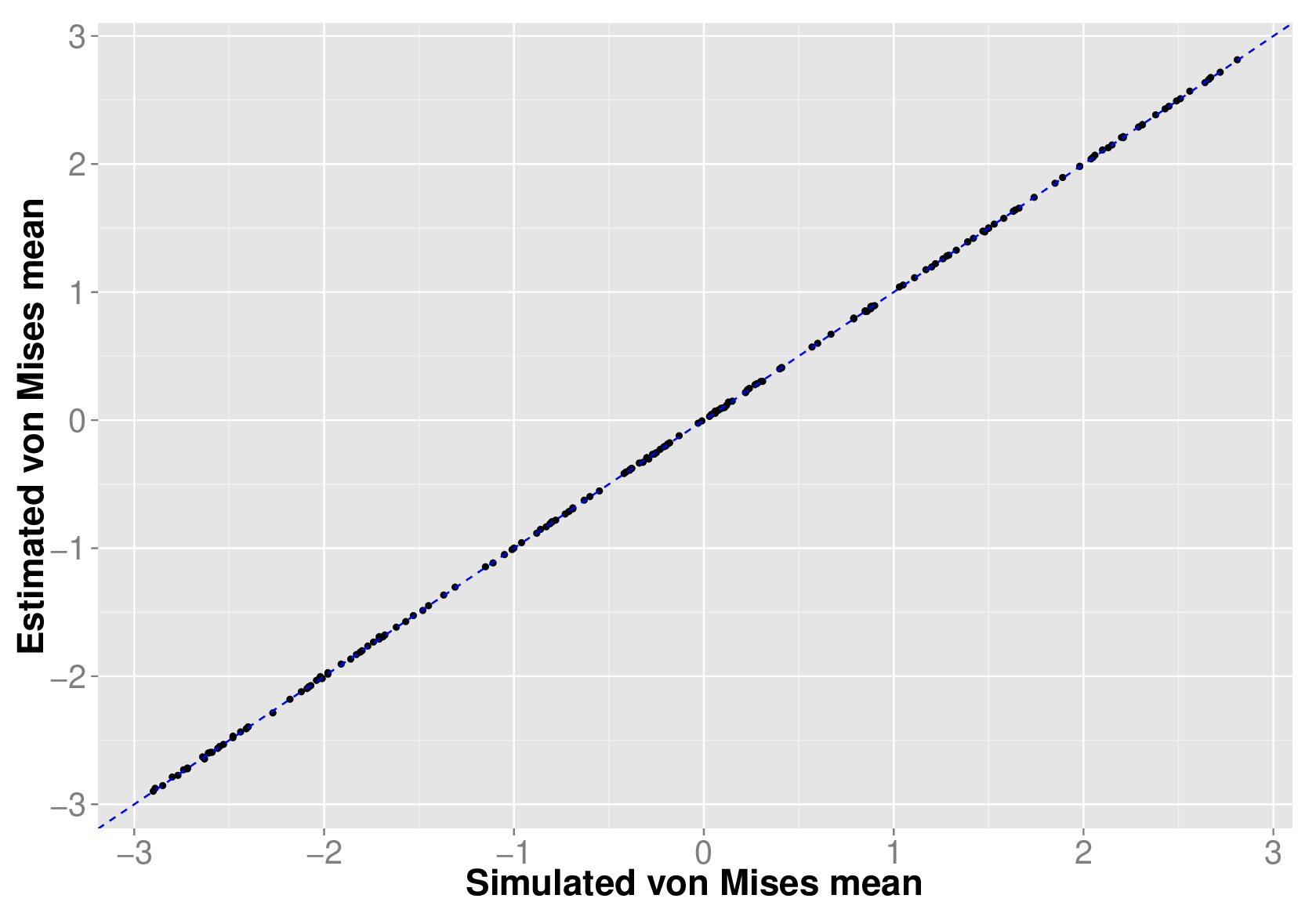}
   \end{minipage}
}
 \subfigure[]{ % FROM THE SUBFIGURE PACKAGE
    \label{fig:d}
    \begin{minipage}[b]{0.5\textwidth}
    \includegraphics[scale=0.3, angle = -90]{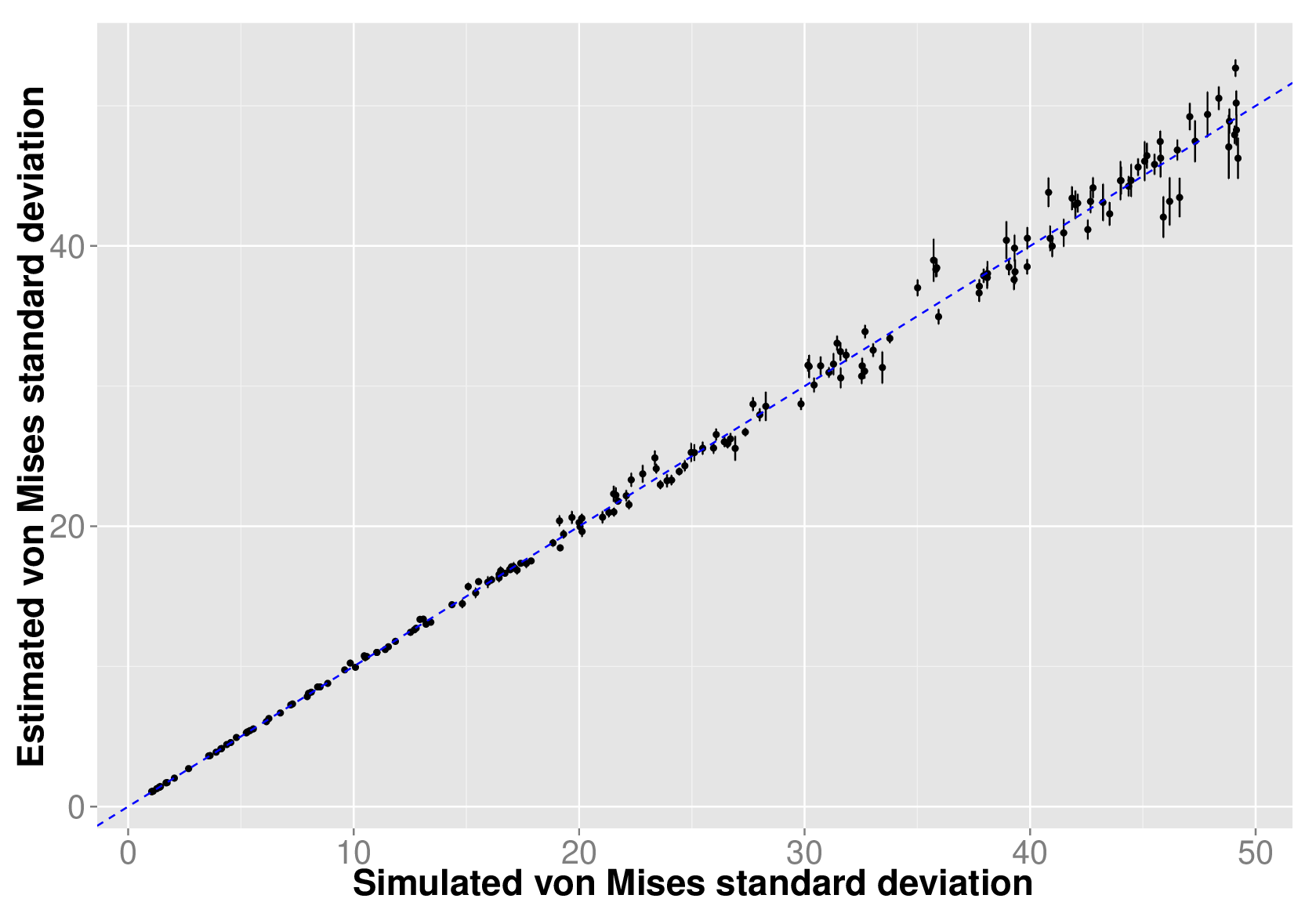}
   \end{minipage}
}

 \subfigure[]{ % FROM THE SUBFIGURE PACKAGE
    \label{fig:d}
    \begin{minipage}[b]{0.5\textwidth}
    \includegraphics[scale=0.3, angle = -90]{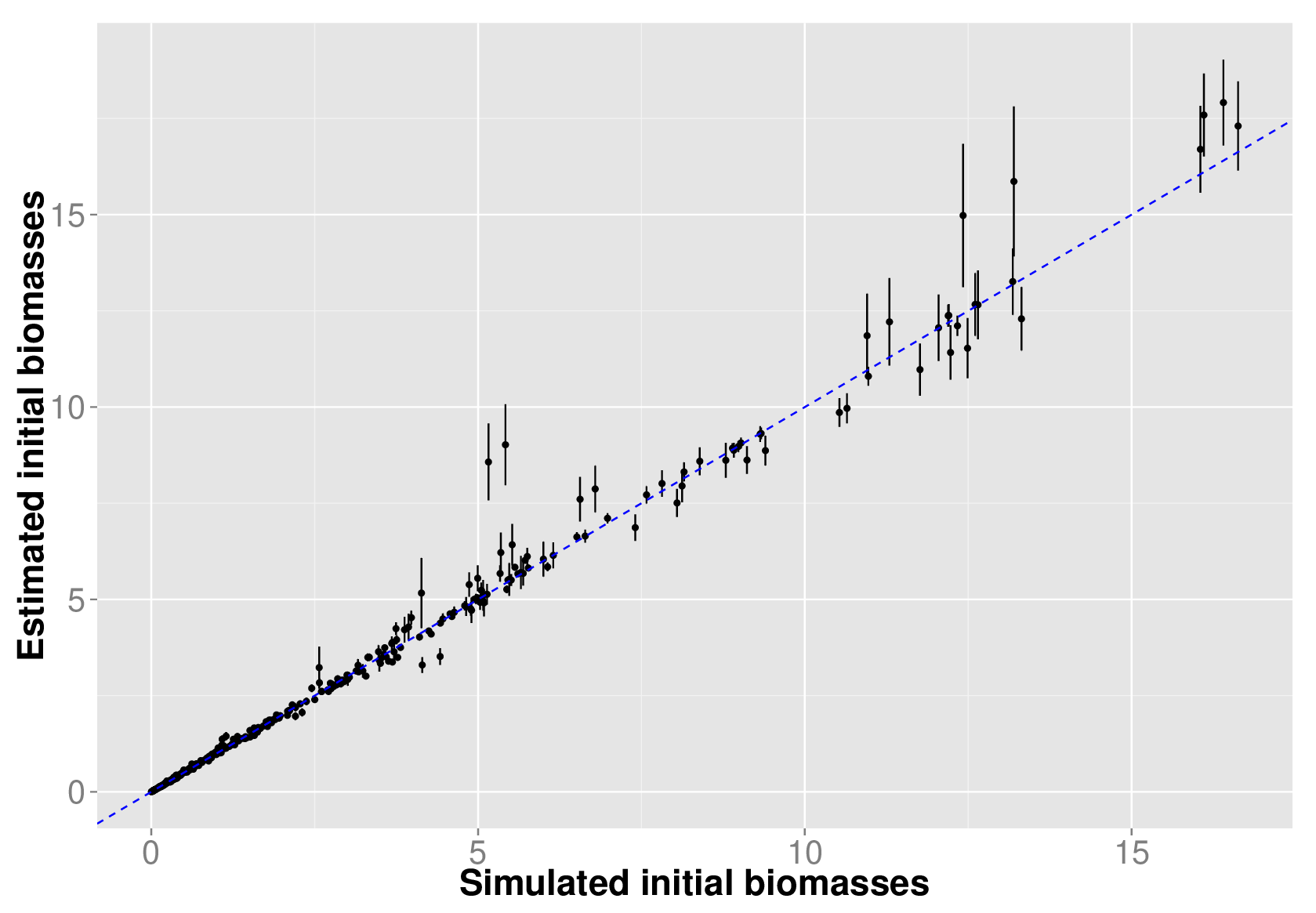}
   \end{minipage}
}

     \caption{Results from 100 simulations testing the capability of the delay difference model coded in C++ to estimate parameters. Each panel compares simulated values (x-axis) to estimated values (y-axis). The vertical bars represent 1 standard deviation. Panel (a) natural mortality; (b) catchability; (c) von Mises distribution mean; (d) von Mises distribution standard deviation and (e) initial biomasses ($B_{1}$ and $B_{2}$).  The dash line represents $y=x$.}

    \label{fig:ResultsOfSimulation}
  \end{figure}

Monte Carlo simulations showed that the implementation of the delay difference in C++ was capable of estimating natural mortality, catchability, von Mises parameters mean and standard deviation and initial biomasses (Fig.~\ref{fig:ResultsOfSimulation}). Simulated values for natural mortality over the range 0.025--0.065 week$^{-1}$ (Tab.~\ref{tab:SimulationParameters.tex}) containing realistic values for tiger prawn could be estimated with less than 31\% discrepancy from simulated values in all cases; less than 15\% discrepancy in 95\% of cases and less than 4\% discrepancy in 50\% of cases. By comparison, catchability estimates were up to 67\% different than their simulated value with 95\% of cases within 18\% discrepancy and 50\% within 2.5\%. \\

\begin{figure}[!ht]
 \subfigure[]{ % FROM THE SUBFIGURE PACKAGE
    \label{fig:CompareOption2and3-HowDeltaAICChangeAsAFunctionOfSimNatMort-1000Simulations}
    \begin{minipage}[b]{0.5\textwidth}
     \includegraphics[scale=0.3, angle = -90]{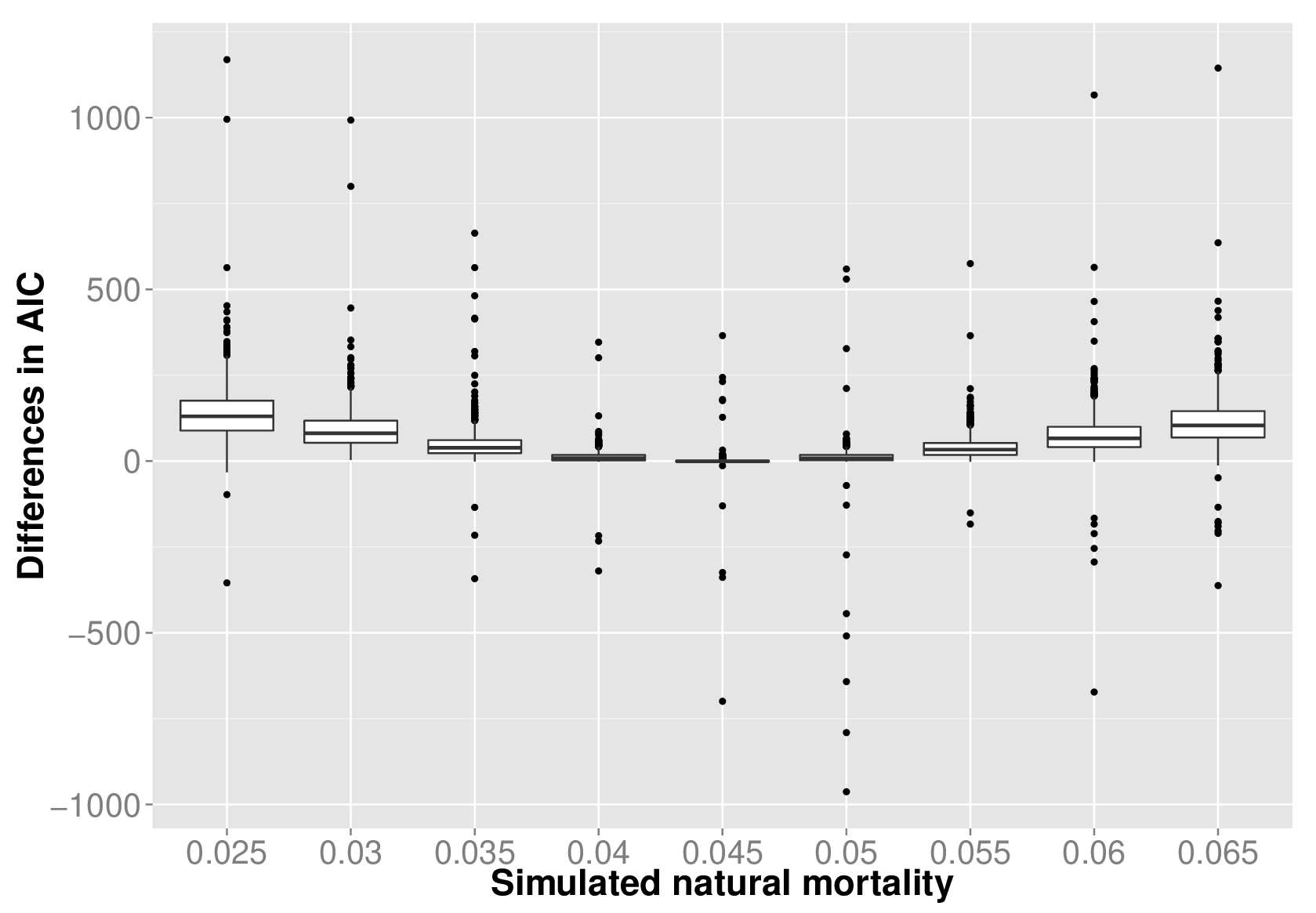}
   \end{minipage}
}
 \subfigure[]{ % FROM THE SUBFIGURE PACKAGE
    \label{fig:CompareOption2and3-HowDiffInCatchabilityChangeAsAFunctionOfSimNatMort-1000Simulations}
    \begin{minipage}[b]{0.5\textwidth}
    \includegraphics[scale=0.3, angle = -90]{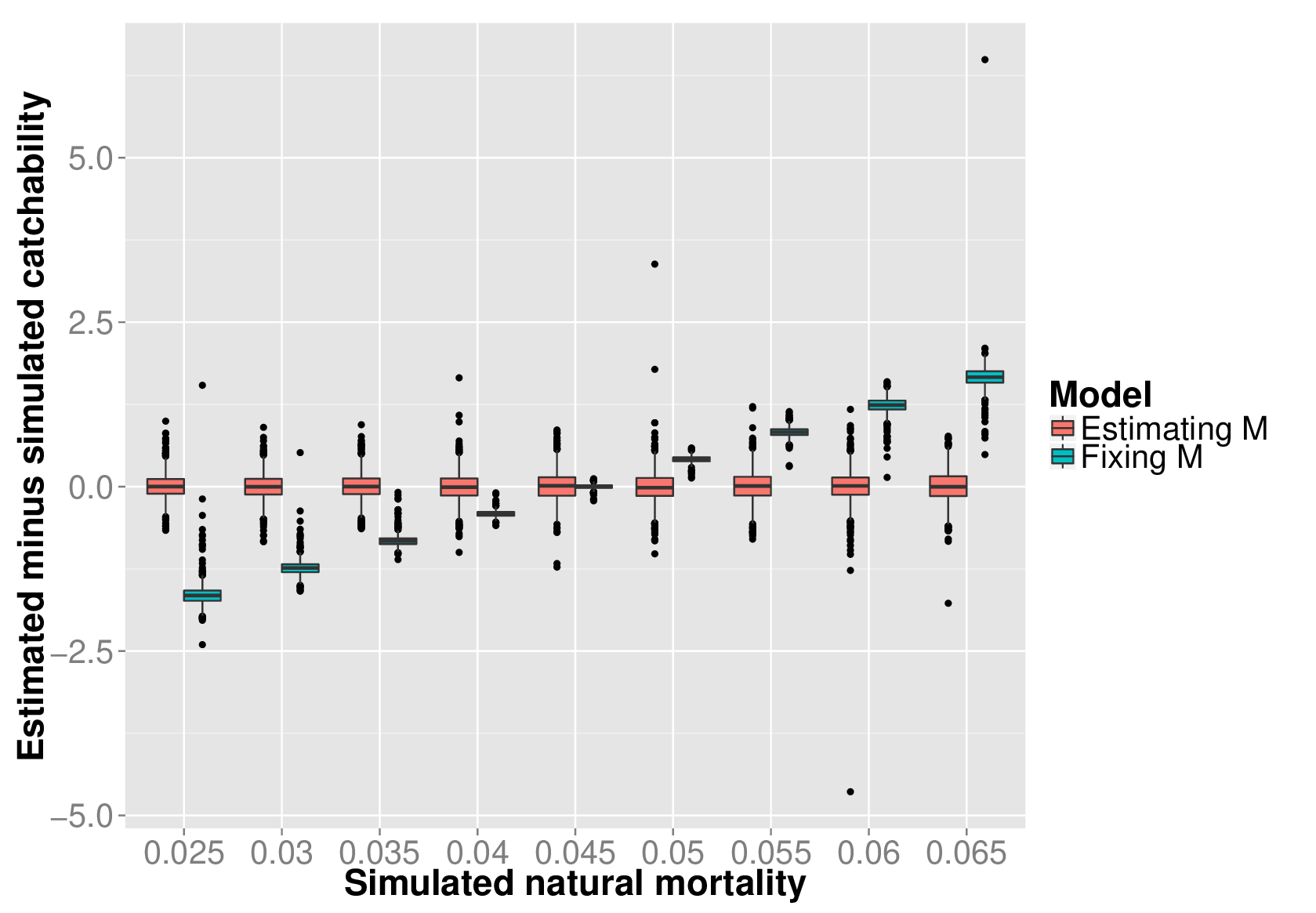}
   \end{minipage}
}
 \subfigure[]{ % FROM THE SUBFIGURE PACKAGE
    \label{fig:Option3-DistributionOfMestimate-1000Simulations.ps} 
    \begin{minipage}[b]{0.5\textwidth}
    \includegraphics[scale=0.3, angle = -90]{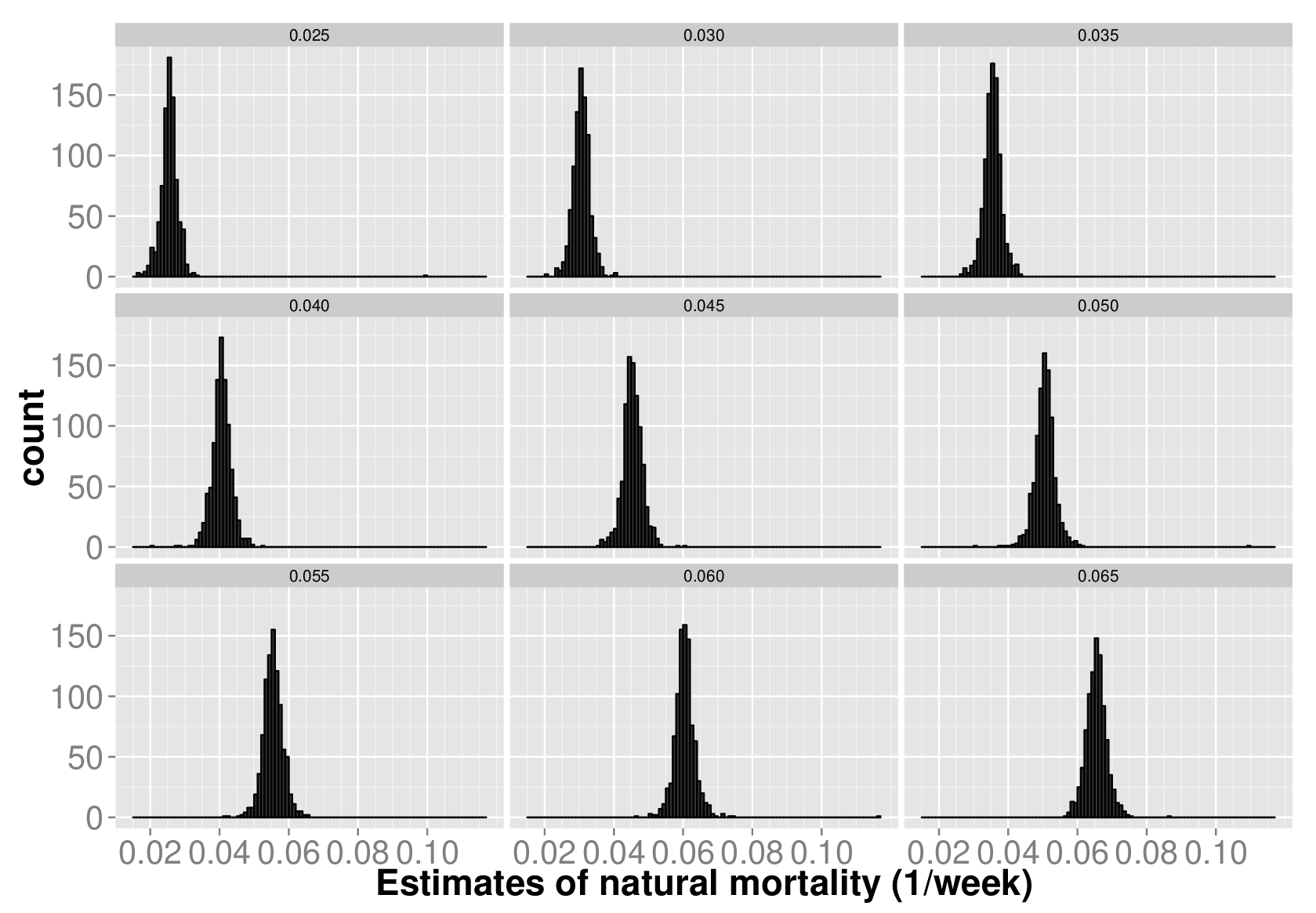}
   \end{minipage}
}
 \subfigure[]{ % FROM THE SUBFIGURE PACKAGE
    \label{fig:IllutrationOfDifferentvonMisesDistributions}
    \begin{minipage}[b]{0.5\textwidth}
    \includegraphics[scale=0.3, angle = -90]{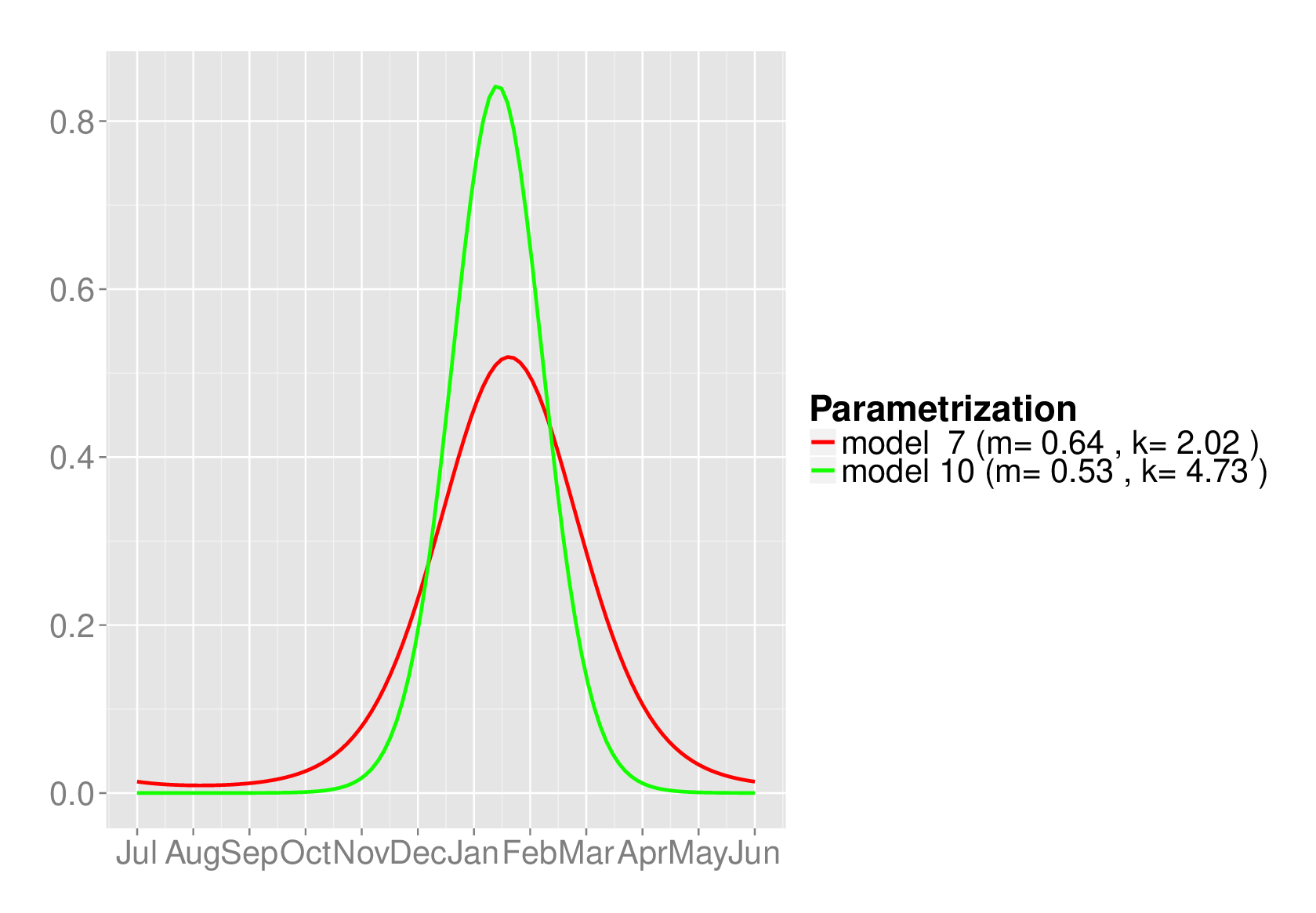}
   \end{minipage}
}

%%  \subfigure[]{ % FROM THE SUBFIGURE PACKAGE
%%     \label{fig:d}
%%     \begin{minipage}[b]{0.5\textwidth}
%%     \includegraphics[scale=0.3, angle = -90]{"/home/mkienzle/mystuff/Work/DEEDI/Biometry services/Fisheries/MoretonBay/TigerPrawn/Analysis/Scripts/MaximumLikelihoodEstimateOfNaturalMortality/TestAbilityToEstimateNaturalMortality/WeeklyTimesteps/Results/Graphics/Option3-EstimateVsSimulate-InitialBiomasses.ps"}
%%    \end{minipage}
%% }
     \caption{Effects of fixing natural mortality in the delay difference model: (a) differences in AIC between model fixing natural mortality and model estimating natural mortality (y-axis) as a function of the simulated value of natural mortality (larger values indicated better fit by estimating natural mortality); (b) effect of fixing natural mortality on catchability estimates and (c) distribution of natural mortality estimates for each value fixed in the simulation and (d) Von Mises distributions using parametrization estimated with different delay difference models.}
    \label{fig:ThreeGraphsOnTheEffectOfFixingNatMort}
  \end{figure}

The AIC was used to measure the benefit of estimating natural mortality: the larger the discrepancy between fixed natural mortality (0.045) and simulated values, the larger the improvement in AIC obtained by fitting the model that estimated natural mortality compared to the one that did not (Fig~\ref{fig:CompareOption2and3-HowDeltaAICChangeAsAFunctionOfSimNatMort-1000Simulations}). \\

Fixing natural mortality in the stock assessment to a specific value influenced estimates of catchability: fixed over-estimates of natural mortality produced under-estimates of catchability and vice-versa (Fig~\ref{fig:CompareOption2and3-HowDiffInCatchabilityChangeAsAFunctionOfSimNatMort-1000Simulations}). The negative correlation between these 2 variables was $\rho=-0.86$.

%   \subsection{Daily temperature estimates and trends} \input{../TowardAnEcologicalStockAssessmentModel/DailyTempEstAndTrends.tex}
   \subsection{Daily temperature estimates and trends} 
\begin{figure}[!ht]
 \subfigure[]{ % FROM THE SUBFIGURE PACKAGE
    \label{fig:CapeMoretonMaxAirTempTSAndSmoothTrend}
    \begin{minipage}[b]{0.5\textwidth}
     \includegraphics[scale=0.3, angle = -90]{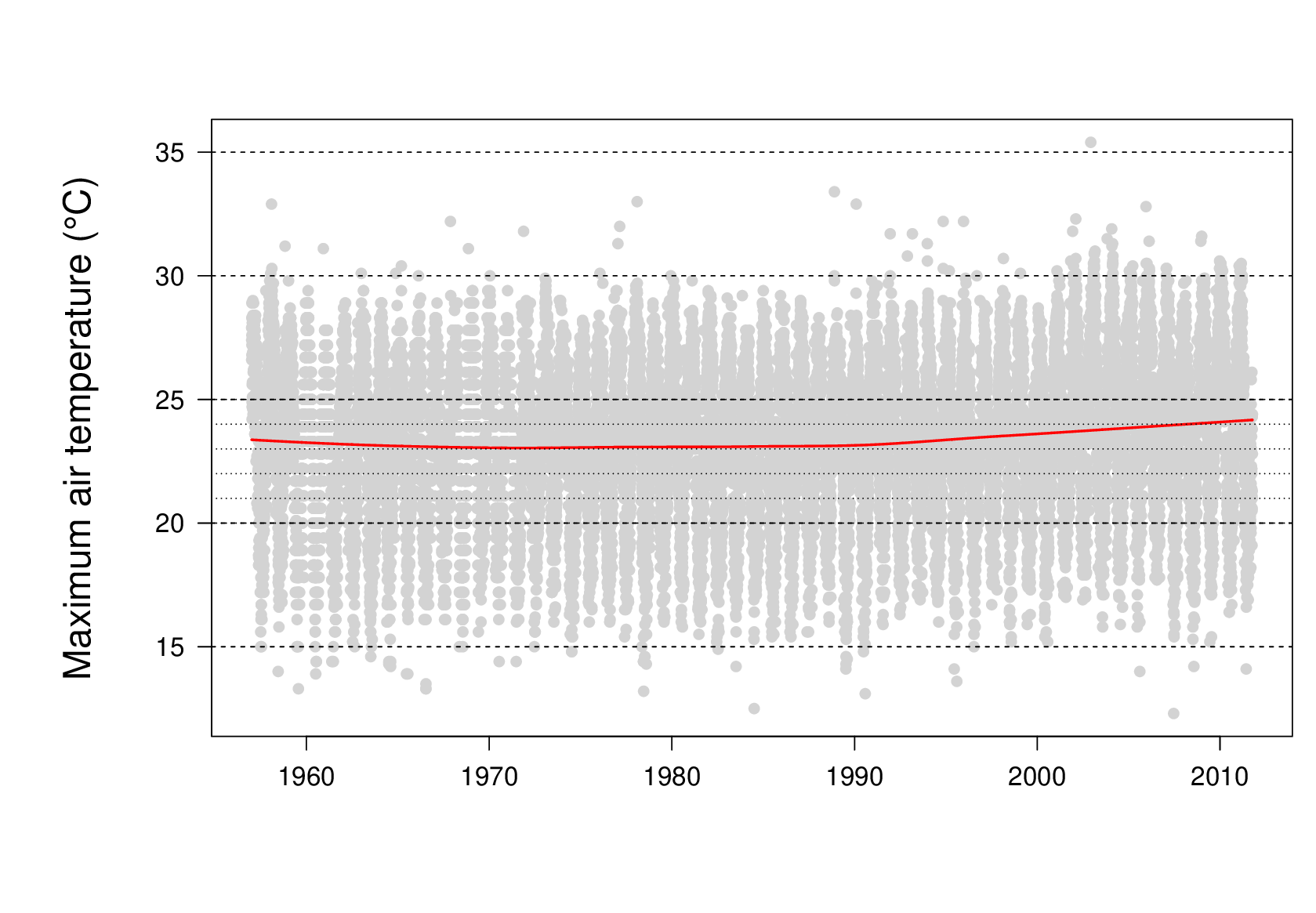}
   \end{minipage}
}
 \subfigure[]{ % FROM THE SUBFIGURE PACKAGE
    \label{fig:LinearRegressionAirSeaLocationE00527}
    \begin{minipage}[b]{0.5\textwidth}
    \includegraphics[scale=0.3, angle = -90]{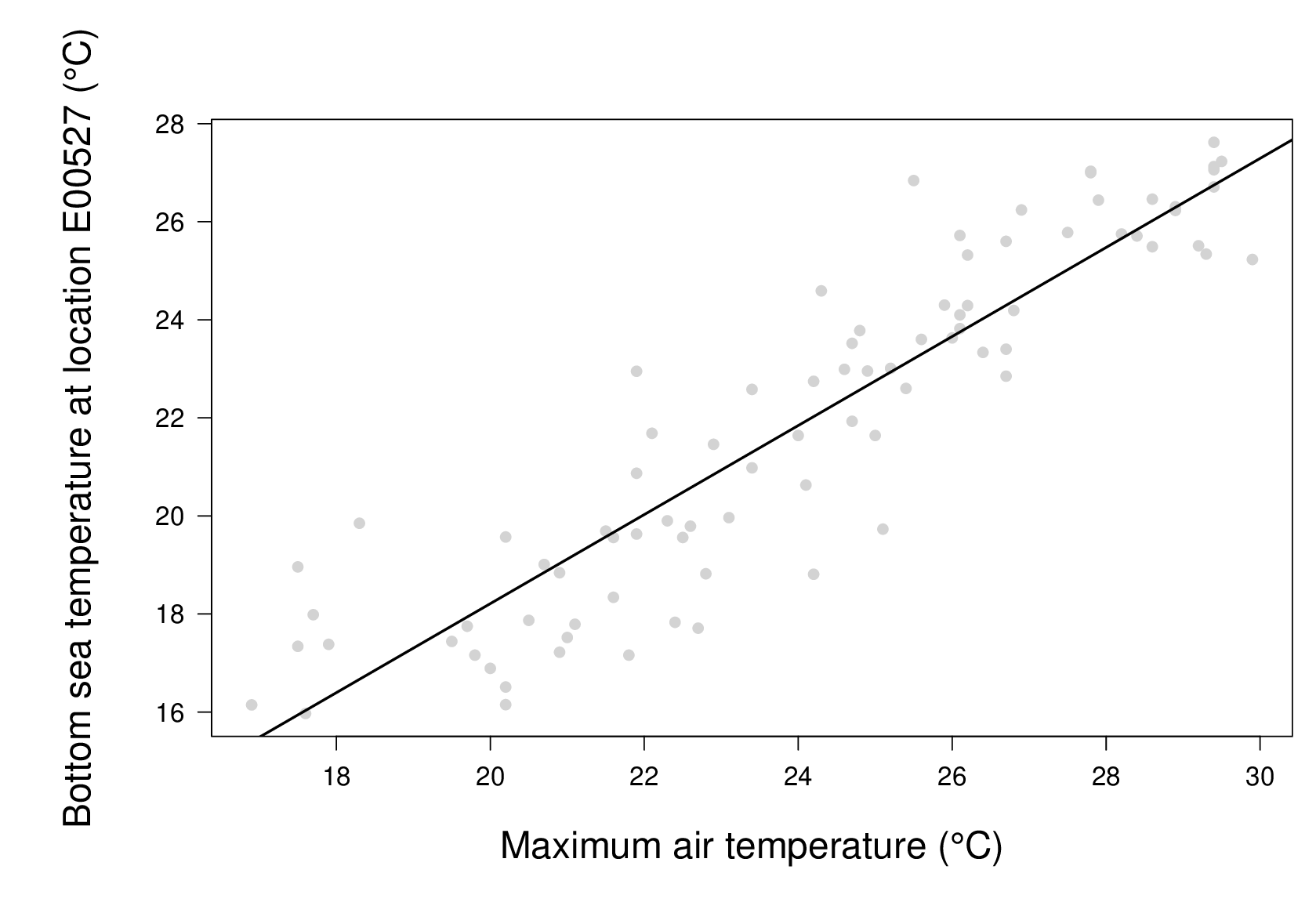}
   \end{minipage}
}
 \subfigure[]{ % FROM THE SUBFIGURE PACKAGE
    \label{fig:EstimatedSeaTemperatureAtLocationE00527}
    \begin{minipage}[b]{0.5\textwidth}
    \includegraphics[scale=0.3, angle = -90]{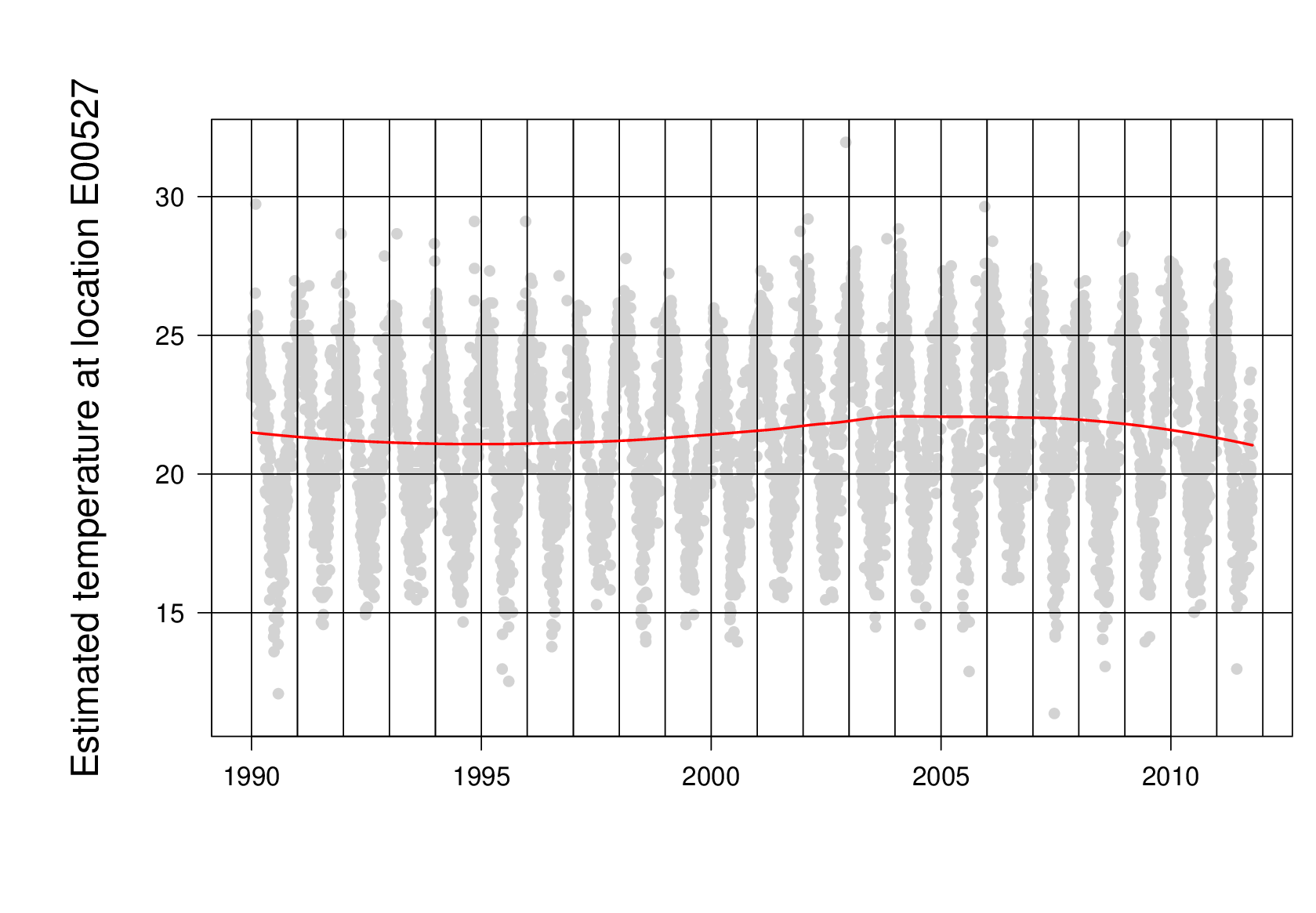}
   \end{minipage}
}
 \subfigure[]{ % FROM THE SUBFIGURE PACKAGE
    \label{fig:LogisticModelOfHillsExpData1985}
    \begin{minipage}[b]{0.5\textwidth}
    \includegraphics[scale=0.3, angle = -90]{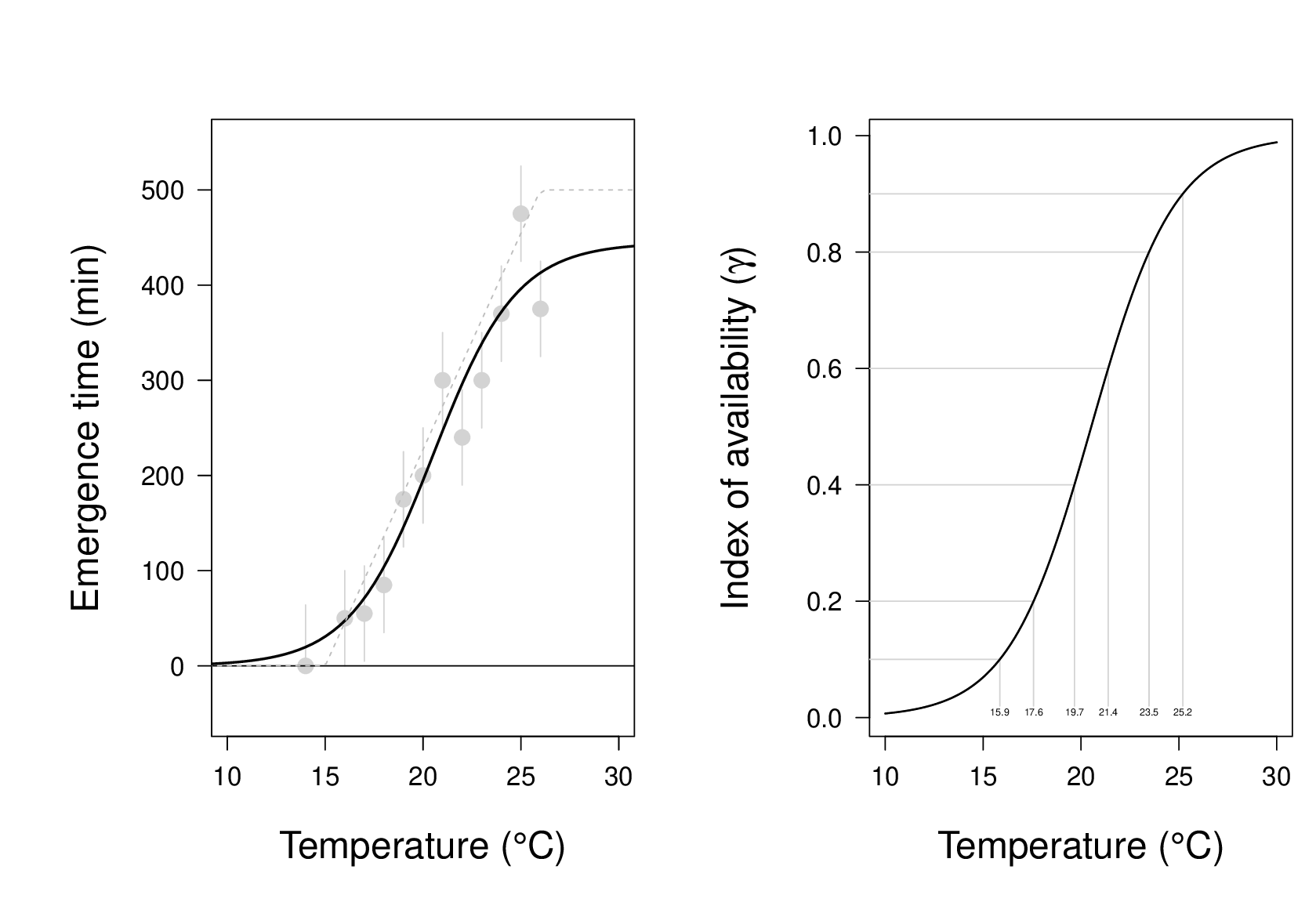}
   \end{minipage}
}
 \subfigure[]{ % FROM THE SUBFIGURE PACKAGE
    \label{fig:EstimatedAvailabilityInRecentYears}
    \begin{minipage}[b]{0.5\textwidth}
    \includegraphics[scale=0.25, angle = -90]{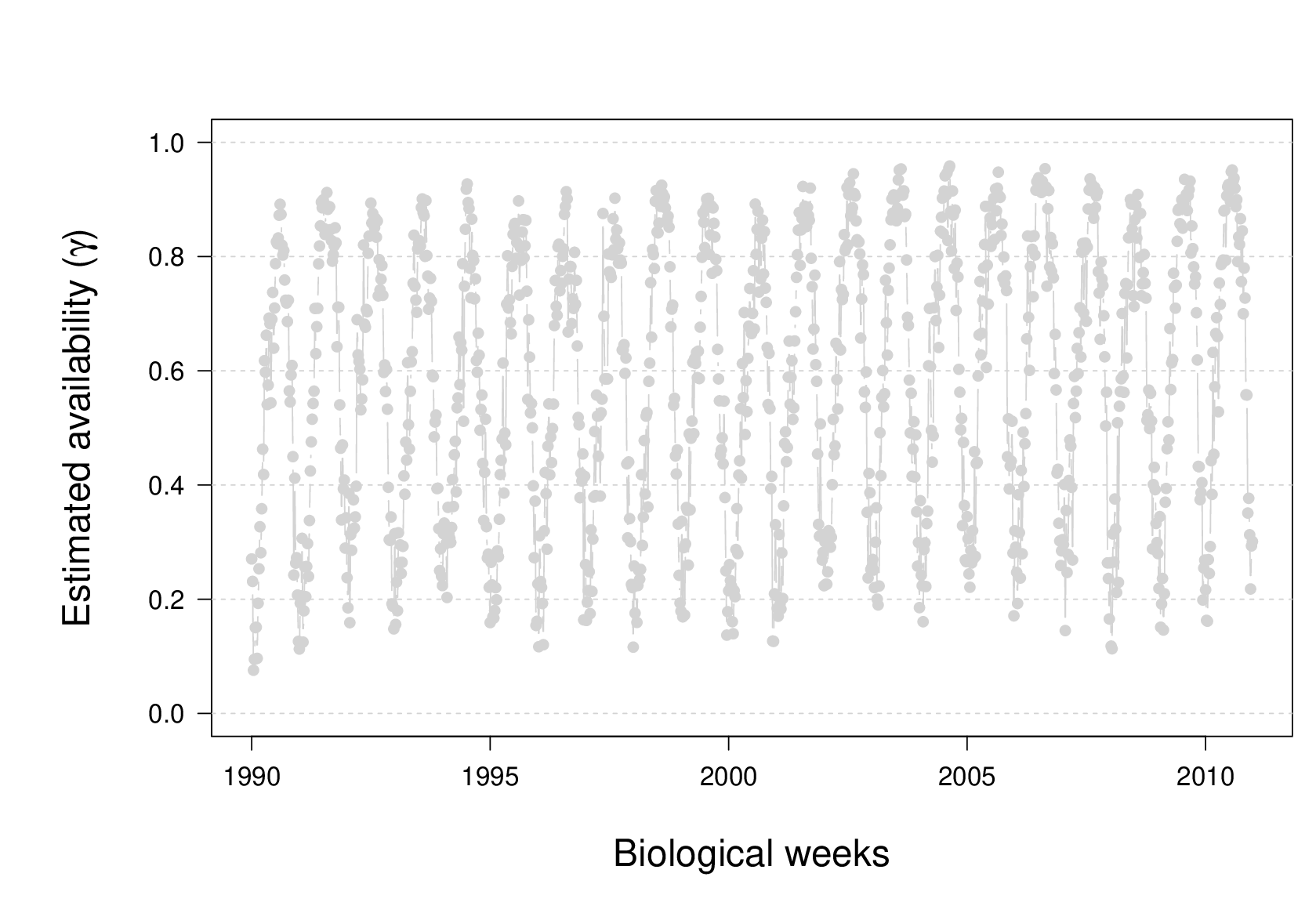}
   \end{minipage}
}
     \caption{Variations of temperature and its effects on availability: (a) time series of maximum air temperature collected by the Australian Bureau of Meteorology at Cape Moreton. The smooth curve represents the trend in the data; (b) linear regression between sea temperature collected at the center of the Bay (station E00527 on y-axis) and maximum daily air temperature recorded at Cape Moreton (x-axis); (c) estimated temperature at location E00527; (d) experimental data on the effect of temperature on the duration of emergence of brown tiger prawn from \cite{hil85a}. The grey vertical bars represent an estimated variability of measurments. The black continuous line represents the fit of a sigmoid function. The grey dotted lines represents the linear model presented by \cite{hil85a} and used in previous stock assessment models \citep{Kienzle2014138} and (e) estimated weekly index of availability.} %(1) using estimated daily temperature measurements and a sigmoid relationship (black dots); (2) using average monthly temperature constant between years and a sigmoid relationship (red dots) and (3) using average monthly temperature constant between years and a linear relationship as in \cite{Kienzle2014138} (green dots).}
    \label{fig:TemperatureData}
  \end{figure}

The maximum daily air temperature, collected by the BOM at Cape Moreton, decreased slightly from 1957 to the mid 1980s and increased in average by 1$^{o}$C from the mid 1980s to the end of 2011 (Fig.~\ref{fig:CapeMoretonMaxAirTempTSAndSmoothTrend}). A linear regression estimated air temperature increased in average by $0.180 \pm 0.014$ $^{o}$C per decade over the entire period. \\

An analysis of variance (ANOVA, Tab.~\ref{tab:anova1}) showed that variables most influencing sea temperature in Moreton Bay were, by decreasing order of importance: time; location and depth. The vertical gradient in temperature was very small because layers of water mixed well in this shallow Bay. The difference in water temperature between surface and deeper layer was $0.0162 \pm 0.012$ $^{o}$C.m$^{-1}$, approx. 0.1$^{o}$C at the average depth of 7 m.\\

A linear regression between 85 matching pairs of sea and air temperature measurements (as a dependent variable) was statistically significant (F-statistic: 354.8 on 1 and 83 DF,  p-value: $< 2.2 \ 10^{-16}$, adjusted R$^{2}=0.81$, Fig.~\ref{fig:LinearRegressionAirSeaLocationE00527}). The intercept did not differ significantly from zero and the estimate of slope was equal to $0.89112 \pm 0.04731$ (Tab.~\ref{tab:ANOVAOfLinearRegressionAirSeaLocationE00527}). The range of estimated daily bottom temperature inside the Bay, at station E00527, between 1988 and 2011 was larger (11.2--32.2$^{o}$C) than ocean water temperature measured off Cape Moreton (20.2--26.1$^{o}$C). Estimated daily sea temperature in the Bay have increased between 1990 and 2010 (Fig.~\ref{fig:EstimatedSeaTemperatureAtLocationE00527}) by an estimated 0.6$^{o}$C using yearly averages in each of the two decades 1990-1999 and 2000-2011.\\

%   \subsection{Model for emergence duration} \input{../TowardAnEcologicalStockAssessmentModel/ResEstimatesOfAvailability.tex}
   \subsection{Model for emergence duration} The sigmoid model fitted well the experimental data on emergence time ($\chi^{2}_{\ 9 \ \rm{df}} = 6.81$, P = 0.34, $\hat{a} = 444.5 \pm 61.11$, $\hat{b} = 9.650 \pm 2.394$ and $\hat{c} = 0.4702 \pm 0.1306$, Fig~\ref{fig:LogisticModelOfHillsExpData1985}). The estimated index of availability ($\gamma(T)$) varied periodically over 12 months, reached its lowest values in southern hemisphere winter and highest in summer (Fig.~\ref{fig:EstimatedAvailabilityInRecentYears}). Its amplitude encompassed differences in availability of 90\% between the warmest summer and the coldest winter.

%   \subsection{Stock assessment model} \input{../TowardAnEcologicalStockAssessmentModel/ResDelayDifferenceModel.tex}
   \subsection{Stock assessment model} Allowing the timing of recruitment to vary from year to year (model 11) provided the largest stock assessment model improvement ($\Delta$AIC = -272.21, Tab.~\ref{tab:ModelComparison}) compared to the second best model (model 10). Mean recruitment was estimated to vary from Jan to March (Fig.~\ref{fig:IllutrationOfDifferentvonMisesDistributions}).\\

Previous work had found that temperature-driven behavioural response that changes catchability seasonally provided large improvements in fitting the delay difference to catch data (models 5--6 and models 3--4, Tab.~\ref{tab:ModelComparison}). The sigmoid model for emergence duration was slightly better than the linear model (compare models 8--3 and 4--3, $\Delta$AIC = -168.97 and $\Delta$AIC = -164.71 resp.). On the other hand, using daily estimates of bottom temperature provided a slight improvement compare to using average monthly temperatures (compare model 9--4 and 9--8, $\Delta$AIC = -49.86 and $\Delta$AIC = -45.6) but the introduction of larger variations in the availability index decreased catchability estimates by nearly half, for both targeted and non-targeted effort. \\

Estimating natural mortality improved further the fit of the delay difference (model 10 in Tab.~\ref{tab:ModelComparison}). In this case, natural mortality was estimated to be equal to $0.035 \pm 0.002$, about 25\% lower than its assumed value of 0.045 wk$^{-1}$ in other model. This additional degrees of freedom in the delay difference also narrowed the distribution of the recruitment (Fig.~\ref{fig:IllutrationOfDifferentvonMisesDistributions}), probably because higher survival meant reduced recruitment magnitude required to explain observed catches. Finally, the model that best fitted the data (model 11) estimated natural mortality equal to $0.031 \pm 0.002$.\\

\section{Discussion} Monte Carlo simulations showed that natural mortality could be estimated within this stock assessment model. The model that best fitted brown tiger prawn data from Moreton Bay estimated natural mortality to be equal to $0.031 \pm 0.002$ week$^{-1}$. This estimate is consistent with that for common banana prawns ({\it Penaeus merguiensis}) obtained by \cite{Zhou25022014} using a depletion method. Both these estimates are approximately 30\% lower than the value of natural mortality commonly used in brown tiger prawn stock assessment (0.045 week$^{-1}$) based on the analysis by \cite{Wang99a} who concluded that natural mortality is at least 0.03 week$^{-1}$ and no more than 0.065 week$^{-1}$. \cite{Zhou25022014}  attributed this discrepancy to differences in behaviour between tiger and banana prawns: the latter forming large aggregations which could improve their chance of surviving predators attacks. An alternative explanation of the difference between natural mortality estimates might be that \cite{Wang99a} used data covering a smaller area and shorter period of time, October to November in a single year, than \cite{Zhou25022014}, evidencing spatial or temporal variations of natural mortality rates of prawns across their habitat due for example to increased predation in spring. Finally, this discrepancy could also have arisen from the difficulty of estimating both catchability and natural mortality simultaneously reported by \cite{Wang99a}, who resorted to fix catchability: in this situation, an under-estimate of catchability could have led to over-estimating natural mortality. At present the information about natural mortality estimates is too sketchy to draw firm conclusions. Further research focused on developing additional independent estimates of natural mortality should help identify the cause(s) of such discrepancy. \\

The conclusion that Moreton Bay brown tiger prawn dataset seems to allow estimation of natural mortality within its stock assessment is difficult to generalize because a big impediment to progressing fisheries science is the availability of good quality data. That Moreton Bay offers, as far as we can see, a reliable logbooks dataset for tiger prawn that is easily fitted by the delay difference model is a meagre consolation against the backdrop of many fisheries which data quality varies from questionable to unusable. Stock assessments scientists often resort to external parameter estimates when data can't provide a reasonable estimate of important aspects of the dynamic of the stock. To a large extent, all assessment models are mis-specified when dealing with real data: they are simplification of a complex reality described only partially by available data. That is, they include many assumptions that we know are false in the real world, but that are needed to make estimation tractable ({\it e.g.}, it is often assumed that neither fishery selectivity nor natural mortality vary from year to year). We cannot avoid mis-specification because the real world is complex and heterogeneous in ways that we can never expect to reproduce in our models \citep{Francis2012LTE}. "Essentially, all models are wrong but some are useful" (George E.P. Box). There are no doubts that the present model is also mis-specified: for example, trawl selectivity have surely changed throughout the period investigated in an unknown (and impossible to know) fashion, yet this implementation of the delay difference model assumes constant knife-edge selectivity fixed throughout the time-series. Ignoring this change in the fishery has certainly had an effect on natural mortality estimates of an unknown magnitude and direction. But the fundamental question raised by this paper is whether to use mortality estimates that ignore all we have learned about the dynamic of this fishery and use, without questioning, a value borrowed from another fishery. Or to allow the stock assessment model to estimate it simultaneously and work at removing as many mis-specifications as we can in order to converge toward a model that provides the best values to parameterize the dynamic of this fishery.\\

Natural mortality is regarded as one of the most influential quantities in fisheries stock assessment and management. The magnitude of natural mortality relates directly to the productivity of the stock, the yields that can be obtained, optimal exploitation rates, and reference points \citep{lee11a}. Since information on total mortality is contained in the catch data, any error in the chosen value of natural mortality, within limits, is compensated by an error in opposite direction on fishing mortality rates: an over-estimate of natural mortality leads to under-estimate catchability and the resulting F will be under-estimated and stock biomasses over-estimated. The stock may appear to be better than it truly is, and fishing mortality appears less of a concern than it should be. For some data poor fisheries, the target fishing mortality is often linked to M ({\it e.g.}, setting F$_{\rm{msy}}$ = M), in such cases over-estimating M will cause stock being over-fished. Reference points estimates, such as maximum sustainable yield, were shown to be robust to natural mortality mis-specifications in a simple age-structure stock assessment \citep{doi:10.1139}, a modest over-estimate of the natural mortality can result in quota recommendations that represent very high real harvest rates. This author recommended to use a conservative (low) estimate of natural mortality for management purposes because underestimating natural mortality rate carry little penalty in terms of long term yield. Investigating the effects of using erroneous natural mortality rate in the delay-difference model on reference points was beyond the scope of this work. But such study would clarify whether using these most recent, smaller, estimates of natural mortality has a significant impact on harvest decisions taken to manage prawn stocks. \\

\section*{Acknowledgements}
This work relied on water temperature measurements graciously provided by Dr. M. Holmes and R. Williams from Queensland Department of Science, Information Technology and the Arts. We are in debt to Dr. J. Robins for her contribution to model developments, in particular her suggestion to look for correlations between air and water temperature. We are grateful to Dr A.J. Courtney for his advice on prawn fisheries.
\clearpage
\newpage
%% Bibliography

%\bibliographystyle{plainnat}
%\bibliography{/home/mkienzle/mystuff/Bibliography/long,/home/mkienzle/mystuff/Bibliography/Biblio}
%\bibliography{long,Biblio}

%hello \cite{aaa}
%% \bibliography{/home/mkienzle/Bibliography/a-JLong}
%% \bibliography{/home/mkienzle/Bibliography/Biblio}
%\bibliographystyle{plain}
%\bibliographystyle{plainnat}
%\bibliographystyle{abbrv}
%\bibliographystyle{apalike}

%% \clearpage
%% \newpage
%% \section*{Figures}

%% %%%%% CPUE and DSITIA sampling station 
%%    \begin{figure}[h!]
%%      \begin{center}
%%        \includegraphics[scale=0.6, angle = 0]{"/home/mkienzle/mystuff/Programs/C++/DelayDifference/Option3/Test the code/WeeklyTimesteps/Results/Graphics/EstimateVsSimulate-NaturalMortality.ps"}
%%        \caption{Natural mortality}
%%        \label{fig:ResultsOfSimulation:NatMort}
%%      \end{center}
%%   \end{figure}

%%%%% CPUE and DSITIA sampling station 

%%%%% Simulations testing of model allowing for natural mortality estimates (Option 3)

%%%%% Three plots showing the effect of using fixed natural mortality rates

\clearpage
\newpage
\section*{Tables}
%\input{Tables/DDModelPar.tex}
% latex table generated in R 2.15.2 by xtable 1.7-0 package
% Mon Feb 25 14:04:24 2013
\begin{table}[ht]
\begin{center}
\begin{tabular}{lll}
  \hline
Parameter & Value & Reference \\
  \hline
  $\rho$  & 0.963      & \cite{grib94a}\\
$w_{k-1}$  & 17.8 grams  & based on the method of \cite{sch85a} \\
$w_{k}$    & 19.5 grams  & based on the method of \cite{sch85a} \\
$M$       & 0.045 week$^{-1}$ & \cite{dichmont2003application}\\
   \hline
\end{tabular}
\caption{Values of parameters fixed in the delay difference model.}
\label{tab:ModelPar}
\end{center}
\end{table}

%\input{Tables/SimulationParameters.tex}
% by marco 17/12/2014

\begin{table}[ht]
\centering
\begin{tabular}{lll}
  \hline
Variable type & Distribution & Parameters \\
  \hline
recruitment               & uniform & min=1e6, max=1e8   \\
von Mises mean            & uniform & min=$-\pi + 0.2$, max = $\pi - 0.2$ \\
von Mises sd              & uniform & min=1, max=50 \\
natural mortality         & uniform & min=0.025, max=0.065 \\
catchability              & uniform & min=1e-4, max=1e-3 \\
fishing effort            & fixed   &  \\
   \hline
\end{tabular}
\caption{Distribution and range of value taken by different variables in simulations.}
\label{tab:SimulationParameters.tex}
\end{table}

%\input{"/home/mkienzle/mystuff/Work/DEEDI/Moreton Bay Prawn Trawl fishery/Analysis/Scripts/InterpolatingTemperatureData/Results/Tables/ANOVA1.tex"}
% latex table generated in R 3.0.2 by xtable 1.7-4 package
% Thu Jan  8 13:58:04 2015
\begin{table}[ht]
\centering
\begin{tabular}{lrrrrr}
  \hline
 & Df & Sum Sq & Mean Sq & F value & Pr($>$F) \\ 
  \hline
Location & 84 & 4366.38 & 51.98 & 42.90 & 0.0000 \\ 
  Depth & 1 & 445.40 & 445.40 & 367.61 & 0.0000 \\ 
  Year:Month & 367 & 606282.41 & 1652.00 & 1363.47 & 0.0000 \\ 
  Residuals & 58585 & 70982.03 & 1.21 &  &  \\ 
   \hline
\end{tabular}
\caption{Analyis of variance of sea water temperature measurements.} 
\label{tab:anova1}
\end{table}

%\input{"/home/mkienzle/mystuff/Work/DEEDI/Moreton Bay Prawn Trawl fishery/Analysis/Scripts/InterpolatingTemperatureData/Results/Tables/ANOVAOfLinearRegressionAirSeaLocationE00527.tex"}
% latex table generated in R 3.0.2 by xtable 1.7-4 package
% Thu Jan  8 14:00:36 2015
\begin{table}[ht]
\centering
\begin{tabular}{rrrrr}
  \hline
 & Estimate & Std. Error & t value & Pr($>$$|$t$|$) \\ 
  \hline
(Intercept) & 0.0486 & 1.0907 & 0.04 & 0.9645 \\ 
  x & 0.9081 & 0.0448 & 20.26 & 0.0000 \\ 
   \hline
\end{tabular}
\caption{Linear regression between sea water temperature in Moreton Bay and maximum air temperature at Cape Moreton.} 
\label{tab:ANOVAOfLinearRegressionAirSeaLocationE00527}
\end{table}

%\input{Tables/ModelComparison.tex}
% latex table generated in R 3.0.2 by xtable 1.7-1 package
% Fri Nov 15 13:22:24 2013
%\begin{table}[ht]
\begin{sidewaystable}[ht]
\centering
\small
\begin{tabular}{lcccccccccc}
  \toprule
  Model & $p$ & AIC & \multicolumn{2}{c}{Catchability} & \multicolumn{2}{c}{Recruitment distribution} & \multicolumn{2}{c}{Initial biomasses} &  \multicolumn{1}{c}{} & \multicolumn{1}{c}{Natural mortality} \\  \cmidrule(r){1-1}
  \cmidrule(r){2-2}
  \cmidrule(r){3-3}
  \cmidrule(r){4-5}
  \cmidrule(r){6-7}
  \cmidrule(r){8-9}
  \cmidrule(r){10-10}
  \cmidrule(r){11-11}
   & & & $q_{1} \ (\times 10^{-4})$ & $q_{2} \ (\times 10^{-4})$ & $\mu$ & $\kappa$ & $B_{1} \ (\times 10^{5})$ & $B_{2} \ (\times 10^{5})$ & $\sigma$ & $M$\\
  \midrule
% latex table generated in R 3.1.1 by xtable 1.7-1 package
% Wed Oct 15 07:12:07 2014
  11 &  49 & 6743.71 & 3.45 $\pm$ 0.532 & 1.424 $\pm$ 0.22 & 0.668 $\pm$ 0.215 & 4.955 $\pm$ 0.591 & 1.012 $\pm$ 0.015 & 0.978 $\pm$ 0.011 & 5.073 $\pm$ 0.114 & 0.031 $\pm$ 0.002 \\ 
   10 &  29 & 7015.92 & 2.245 $\pm$ 0.377 & 0.839 $\pm$ 0.149 & 0.525 $\pm$ 0.03 & 4.734 $\pm$ 0.586 & 2.173 $\pm$ 0.682 & 2.038 $\pm$ 0.615 & 5.853 $\pm$ 0.127 & 0.035 $\pm$ 0.002 \\ 
    9 &  28 & 7040.14 & 2.614 $\pm$ 0.801 & 1.052 $\pm$ 0.285 & 0.61 $\pm$ 0.061 & 2.819 $\pm$ 0.391 & 2.078 $\pm$ 0.82 & 1.923 $\pm$ 0.741 & 5.924 $\pm$ 0.132 &  \\ 
    7 &  28 & 7071.04 & 4.089 $\pm$ 0.425 & 1.858 $\pm$ 0.255 & 0.642 $\pm$ 0.036 & 2.022 $\pm$ 0.088 & 0.457 $\pm$ 0.011 & 0.447 $\pm$ 0.012 & 6.008 $\pm$ 0.13 &  \\ 
    8 &  28 & 7085.74 & 4.761 $\pm$ 0.485 & 1.898 $\pm$ 0.247 & 0.626 $\pm$ 0.035 & 2.053 $\pm$ 0.09 & 0.375 $\pm$ 0.15 & 0.372 $\pm$ 0.131 & 6.049 $\pm$ 0.133 &  \\ 
    4 &  28 & 7090.00 & 4.164 $\pm$ 0.44 & 1.713 $\pm$ 0.236 & 0.649 $\pm$ 0.036 & 2.047 $\pm$ 0.09 & 0.451 $\pm$ 0.012 & 0.444 $\pm$ 0.013 & 6.06 $\pm$ 0.133 &  \\ 
    6 &  27 & 7162.90 & 1.612 $\pm$ 0.215 &  & 0.513 $\pm$ 0.022 & 5.29 $\pm$ 0.582 & 1.025 $\pm$ 0.342 & 0.979 $\pm$ 0.301 & 6.272 $\pm$ 0.132 &  \\ 
    3 &  28 & 7254.71 & 4.268 $\pm$ 0.357 & 1.888 $\pm$ 0.222 & 0.499 $\pm$ 0.032 & 2.358 $\pm$ 0.128 & 0.148 $\pm$ 0.079 & 0.163 $\pm$ 0.068 & 6.535 $\pm$ 0.144 &  \\ 
    2 &  28 & 7265.15 & 5.034 $\pm$ 0.42 & 1.883 $\pm$ 0.228 & 0.471 $\pm$ 0.032 & 2.254 $\pm$ 0.118 & 0.128 $\pm$ 0.068 & 0.142 $\pm$ 0.059 & 6.566 $\pm$ 0.141 &  \\ 
    5 &  27 & 7332.61 & 3.243 $\pm$ 0.283 &  & 0.513 $\pm$ 0.026 & 3.386 $\pm$ 0.269 & 0.182 $\pm$ 0.09 & 0.189 $\pm$ 0.077 & 6.778 $\pm$ 0.147 &  \\ 
    1 &  27 & 7369.77 & 3.463 $\pm$ 0.299 &  & 0.486 $\pm$ 0.023 & 3.612 $\pm$ 0.31 & 0.17 $\pm$ 0.087 & 0.177 $\pm$ 0.075 & 6.894 $\pm$ 0.154 &  \\ 
  
   \hline
\end{tabular}
\caption{Comparison between several models fitted to catch using a negative log-likelihood function. Results were ordered by increasing values of Akaike Information Criterion (AIC) from top to bottom.} 
\label{tab:ModelComparison}
\end{sidewaystable}
%\end{table}

%\input{./Tables/Calendar.tex}
%%\input{../../Analysis/Scripts/InterpolatingTemperatureData/Results/Tables/ANOVA1.tex}

\end{document}